\documentclass[showpacs,apl,twocolumn,floatfix]{revtex4-1}

\usepackage{graphicx,float,amsmath}
\usepackage{ mathrsfs }
\usepackage{natbib}

\begin{document}
\date{\today}

\title{Anomalous temperature dependence of photoelectron charge and spin mobilities in p$^+$-GaAs.}

\author{F. Cadiz}
\email{fabian.cadiz@polytechnique.edu}
\affiliation{Physique de la mati\`ere condens\'ee, Ecole Polytechnique, CNRS, 91128 Palaiseau, France}

\author{D. Paget}
\affiliation{Physique de la mati\`ere condens\'ee, Ecole Polytechnique, CNRS, 91128 Palaiseau, France}

\author{A.C.H. Rowe}
\affiliation{Physique de la mati\`ere condens\'ee, Ecole Polytechnique, CNRS, 91128 Palaiseau, France}

\author{E. Peytavit}
\affiliation{Institut d'Electronique, de Micro\'electronique et de Nanotechnologie (IEMN), University of Lille, CNRS, Avenue Poincar\'e, Cit\'e Scientifique, 59652 Villeneuve d'Ascq, France}

\author{S. Arscott}
\affiliation{Institut d'Electronique, de Micro\'electronique et de Nanotechnologie (IEMN), University of Lille, CNRS, Avenue Poincar\'e, Cit\'e Scientifique, 59652 Villeneuve d'Ascq, France}

\begin{abstract}
The effect of an electric field on the spatial charge and spin profiles of photoelectrons in p$^+$-GaAs  is studied as a function of lattice and electron temperature. The charge and spin mobilities of photoelectrons are equal in all conditions and exhibit the well known increase as the temperature is lowered. It is shown that this is related mainly to the electron statistics rather than the majority hole statistics. This finding suggests that current theoretical models based on degeneracy of majority carriers cannot fully explain the observed temperature dependence of minority carrier mobility.
 
\end{abstract}
\maketitle

Understanding the mechanisms which limit the minority carrier charge mobility $\mu_e$ and spin mobility $\mu_s$ in semiconductors is necessary for the correct design of bipolar charge and spin devices. The limiting mechanisms are revealed in studies of the temperature dependence of the average momentum relaxation time $\langle\tau_m\rangle=\mu_e m^*/q$ where $q$ is the absolute value of the electronic charge and $m^*$ is the effective mass. Here $\tau_m = \tau_m(\varepsilon) = a(T)\varepsilon^p$ is dependent on the electron's kinetic energy, $\varepsilon$, and the notation $\langle\tau_m\rangle$ indicates an average over all electrons. The exponent $p$ depends on the process which limits the mobility \cite{smith1978} and this energy dependence results in a temperature dependence of $\tau_m$ given, in the nondegenerate case, by 

\begin{equation}
\tau_m \propto a(T).T^{-p}  
\label{tau}
\end{equation}

\noindent
where $k_B$ is Boltzmann's constant and $T$ is the temperature.\ 

The electron mobility in p-type GaAs features a completely different temperature dependence than that of majority electrons in n-GaAs for a similar doping level \cite{beyzavi1991,luber2006,schultes2013}. A number of possible reasons for this have been discussed in the literature, including carrier freezeout at high hole concentrations \cite{kim1997,lovejoy1995}, screening of ionized impurities \cite{walukiewicz1979} and increasing hole degeneracy as temperature is lowered \cite{kaneto1993}. The correct determination of $p$, and thus the role and importance of each of these phenomena, remains uncertain since it is not clear whether the lattice, hole or electron temperature should be used in Eq. \ref{tau}. 

\begin{figure}[tbp]
\includegraphics[clip,width=8.5 cm] {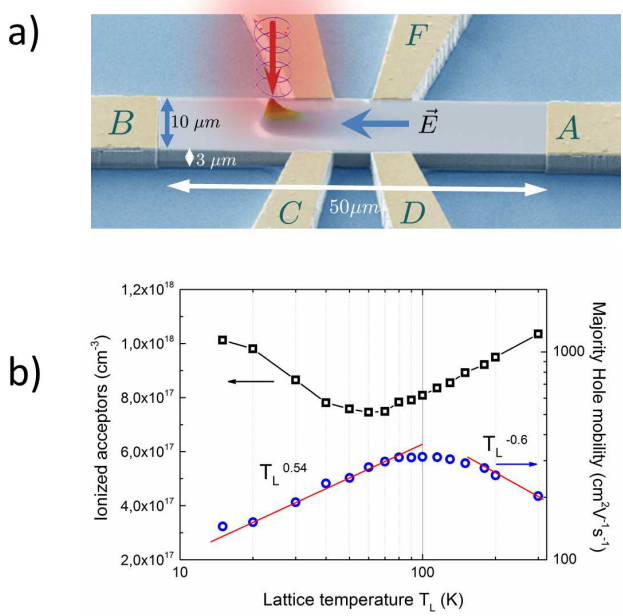}
\caption{(a) SEM image of the p-GaAs hall bar used to study both minority and majority carrier transport. Also shown is a 3D representation of a measured electron density profile when an electric field is applied. (b) Concentration of ionized acceptors and majority hole mobility as a function of lattice temperature, as found from resistivity and Hall effect measurements.}
\label{Fig01}
\end{figure}

Here we experimentally measure $p$ in a 3 $\mu$m thick, carbon doped p-GaAs active layer ($N_A= 10^{18}\;\mbox{cm}^{-3}$) and demonstrate the crucial role of the photoelectron temperature, $T_e$ in Eq. \ref{tau}. The interface between the active layer and the S.I. GaAs substrate is a 50 nm thick GaInP epilayer that confines photoelectrons to the active layer and ensures that recombination at the interface is negligible. As shown in Fig. \ref{Fig01}(a), a Hall bar is photolithographically etched into the active layer so that the lattice temperature ($T_L$) dependence of the majority hole concentration and mobility can be measured as shown in Fig. \ref{Fig01}(b). The density of ionized acceptors is only weakly temperature dependent for this doping level since the impurity band and the valence band are merged into a continuum of states \cite{kim1997,lovejoy1995}. The hole mobility at room temperature, $\mu_h=202 \;  \mbox{cm}^2\mbox{V}^{-1}\mbox{s}^{-1}$, as well as its $\sim T_L^{1/2}$ temperature dependence below 100 K are in good agreement with previous reports on similarly doped GaAs \cite{colomb1992,harmon1993,lovejoy1995}.

\begin{figure}[htbp]
\includegraphics[clip,width=8.5 cm] {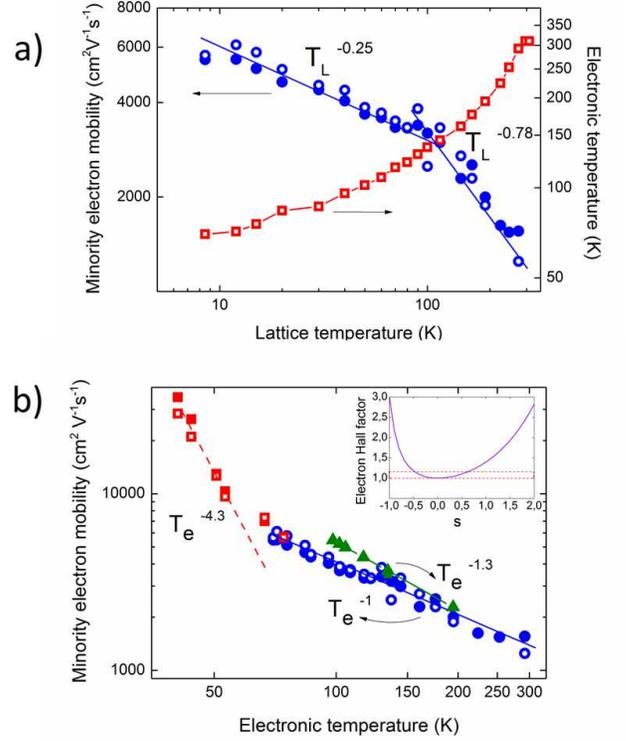}
\caption{(a) Normalized luminescence spectra for different lattice temperatures. Fits of the high energy tail of the spectra (shown by thicker lines) are used to estimate the electronic temperature. (b) Measured electronic temperature as a function of applied electric field for different values of the electron temperature at zero electric field. The increase of $T_e$ a function of $E$ is well explained by simple energy balance considerations which assume an energy-independent energy-relaxation time of $1.5$ ps (dotted line) and $1.2$ ps (solid line) for each case.}
\label{Fig02}
\end{figure}

Spin-polarized photoelectrons were generated by a tightly-focussed circularly-polarized CW laser excitation (1/e half width of $0.6\; \mu$m, energy $1.58$ eV) in a setup described elsewhere \cite{favorskiy2010}. A maximum excitation power of 0.01 mW produces a non degenerate photoelectron concentration of $\sim 5 \times 10^{14}\;\mbox{cm}^{-3}$ in the steady state. $T_e$ was monitored from the high energy tail of the spatially homogenous luminescence spectrum, which is obtained using a multimode optical fiber placed in the image plane. As shown in Fig. \ref{Fig02}(a), taken at a low electric field of $E=200$ V/cm, $T_e$ significantly differs from $T_L$. Furthermore, Fig. \ref{Fig02}(b) shows that without causing intervalley transitions \cite{furuta1990,okamoto2014}, $T_e$ can be increased by applying higher electric fields. The lines in Fig. \ref{Fig02}(b) are predictions based on a simple balance equation $3/2 k_B [T_e(E) - T_e(0)] = q v_d E \tau_E$ for the power delivered to the electron gas by the electric field. Here, $T_e(0)$ is the electronic temperature at zero electric field, $v_d=\mu_e E$ is the drift velocity and $\tau_E$ the energy relaxation time. The data is well explained by an energy independent energy relaxation time of $\tau_E=1.2$ ps and $\tau_E=1.5$ ps for initial electron temperatures of $T_e(0)=40$ K and $T_e(0)= 92$ K, respectively. These values are an order of magnitude larger than values measured at $300$ K \cite{furuta1990}, consistent with a significant decrease of the energy relaxation rate at lower temperatures due to less efficient phonon scattering. Note that under similar electric fields, resistivity measurements in the dark show that the hole temperature ($T_h$) remains unchanged and that $T_h \approx T_L$. This is not surprising since the photogenerated hole concentration is much smaller than $N_A$.

\begin{figure}[htbp]
\includegraphics[clip,width=8 cm] {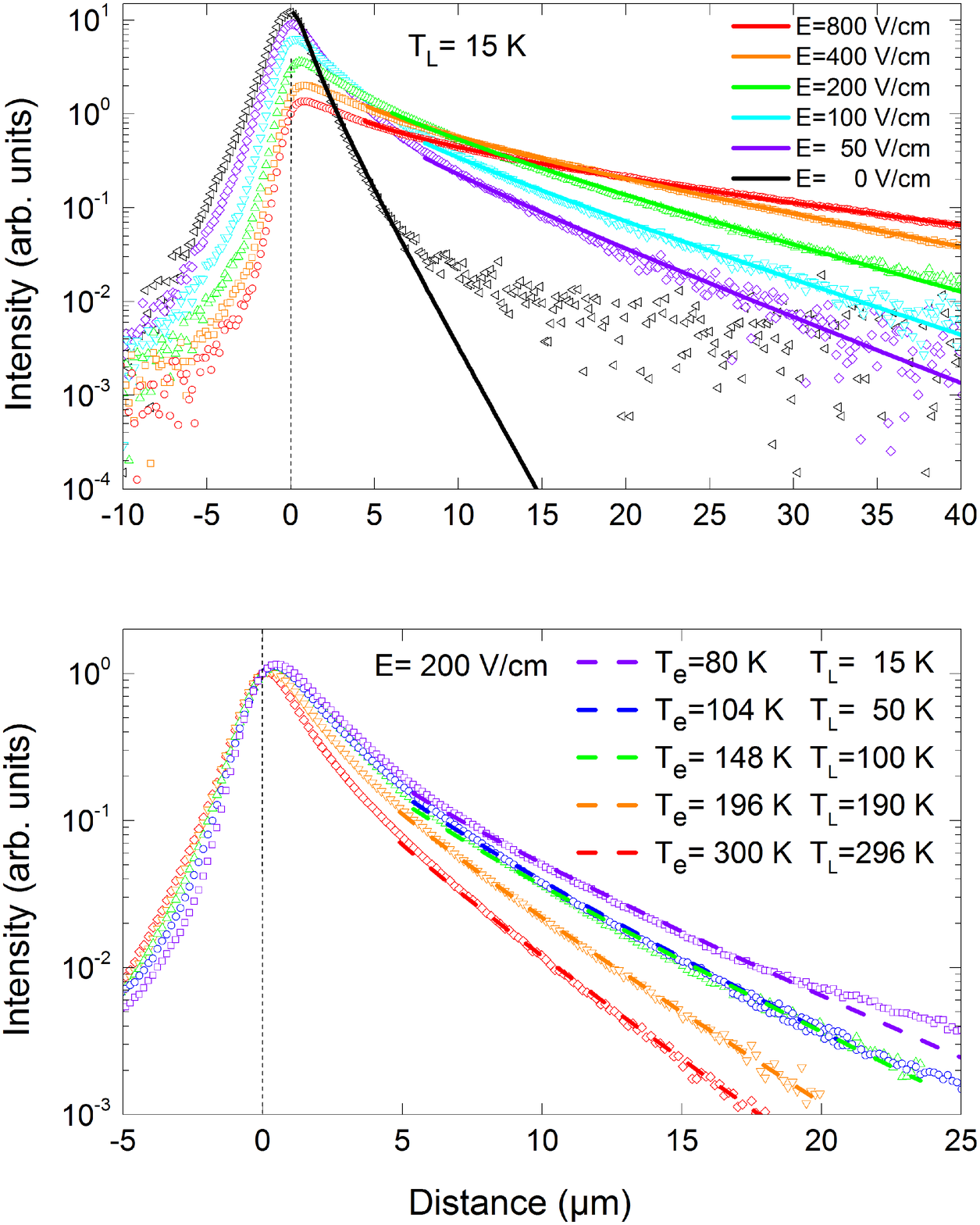}
\caption{The top panel shows the charge density profiles at $T_L = 15$ K for selected values of the electric field. Dotted lines are fits obtained with Eq. \ref{charge} that give the $\mu_e \tau_e$ product for the charge distribution. The bottom panel shows the same profiles at different lattice (and electronic) temperatures when an electric field of 200 V/cm is applied.}
\label{Fig03}
\end{figure}

The minority carrier mobility $\mu_e$ is determined by imaging the spatial dependence of the luminescence, which is proportional to the electron concentration $n$ \cite{favorskiy2010}. Fig. \ref{Fig03} shows sections of these images along the direction of the electric field for $T_L =$ 15 K (see movie attached as an ancillary file). Drift of the electrons in the applied electric field leads to a significant change of the profiles which are well approximated by the 2-dimensional diffusion result \cite{luber2006}
\begin{equation}
n(x) \propto e^{\mu_e \tau_e E x} K_0\left[\frac{\sqrt{(\mu_e \tau_e E)^2 + 4D_e \tau_e}}{2D_e \tau_e}\; x \right],
\label{charge}
\end{equation}
where $D_e$ and $\tau_e$ are the electron diffusion constant and lifetime, respectively, and $K_0$ is a modified Bessel function of the second kind. In a nondegenerate electron gas the Einstein relation, $D_e = [k_BT_e/q] \mu_e $, is valid so that the only fitting parameter in Eq. \ref{charge} is the $\mu_e \tau_e$ product. An independent time resolved photoluminescence characterization of the sample was made in order to measure $\tau_e$ as a function of electron temperature \cite{cadiz2014}, so that a fit of Eq. \ref{charge} to the luminesence intensity profile yields an estimate for $\mu_e$. At $T_L =$ 300 K, $\mu_e= 1560\;\mbox{cm}^2\mbox{V}^{-1}\mbox{s}^{-1}$, in excellent agreement with the theoretical value of $1643\;  \mbox{cm}^2\mbox{V}^{-1}\mbox{s}^{-1}$ at similar doping predicted by Bennett \cite{bennet2002} and with the existing experimental data \cite{harmon1993,colomb1992,beyzavi1991, ahrenkiel1987}. The spin mobility $\mu_s$ was measured using a similar approach by placing a circularly-polarized excitation and a quarter wave plate followed by a linear polarizer at the reception, thus yielding the profile of the spin concentration $s=n_+ - n_-$, where $n_{\pm}$  is the concentration of electrons of spin $\pm$, with a quantization axis along the direction of light excitation \cite{favorskiy2010}. The equation for $s$ is similar to Eq. \ref{charge} where $D_e$ and $\tau_e$ are replaced by their spin counterparts $D_s$ and $\tau_s$. Using the predetermined value of the spin lifetime, $\tau_s$ \cite{cadiz2014}, an estimation of $\mu_s$ is possible.

\begin{figure}[htbp]
\includegraphics[clip,width=8.5 cm] {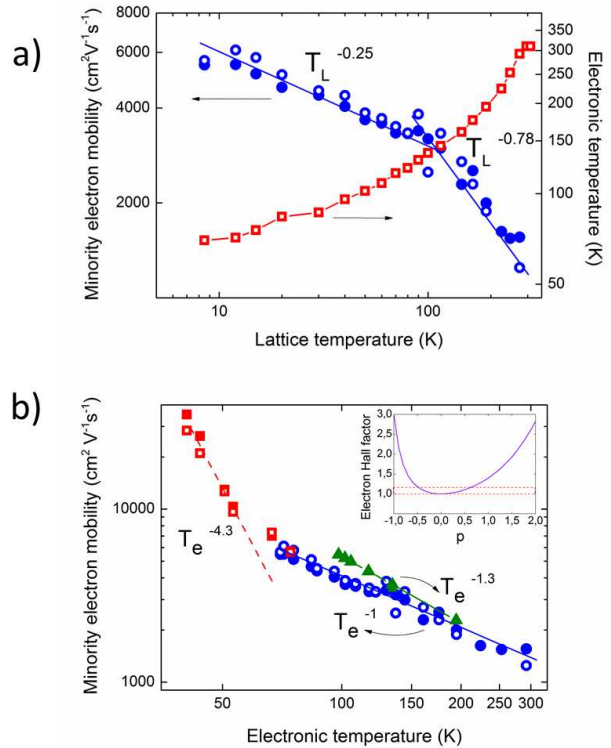}
\caption{(a) The measured electron (solid circles) and spin (open circles) mobilities for an electric field of $200$ V/cm as a function of $T_L$. Also shown (open squares) is the variation of $T_e$. (b) The fixed field, variable $T_L$ mobility data plotted against $T_e$ shows a $1/T_e$ variation at high temperatures, a variation which is also found by varying $T_e$ using the electric field only (i.e. fixed $T_L$, variable field). The solid triangles correspond to $T_L \approx 30-40$ K and an electricl field varying between $200-1300$ V/cm. The solid and open squares correspond to $T_L = 15$ K and a variable electric field ($50-800$ V/cm). At low $T_e$ a dramatic increase of the mobility is observed, with a $1/T_e^{4.3}$ dependence. The inset of the bottom panel shows the expected dependence of $r_H$ (Eq. \ref{rh}) on the index $p$ in Eq. \ref{tau}.}
\label{Fig05}
\end{figure}

The results are summarized in Fig. \ref{Fig05}. The charge and spin mobilities as a function of $T_L$ for a fixed electric field of 200 V/cm are shown in Fig. \ref{Fig05}(a). It is first found that $\mu_e$ and $\mu_s$ equal within experimental error at all temperatures, thus verifying predictions for a nondegenerate electron gas in the absence of spin-dependent momentum relaxation mechanisms \cite{flatte2006}. For $T_L > 100$ K, $\mu_e$ varies as $T_l^{-0.78}$ close to the $1/T_L$ behaviour previously reported in this temperature range \cite{beyzavi1991}. This exponent is reduced to -0.25 for $T_L <$ 100 K. Fig. \ref{Fig05}(a) also shows a linear variation of $T_e$ with $T_L$ above 100 K followed by a weaker variation at lower lattice temperatures. 

The fixed field, variable $T_L$ mobility data are again plotted in Fig. \ref{Fig05}(b) (full and open circles), this time as a function of $T_e$. Above 100 K, a clear $1/T_e$ dependence is again observed, similar to that reported elsewhere \cite{beyzavi1991}. This suggests that what really determines minority carrier mobility is their own statistics, imposed by the light excitation energy and power, rather than that of majority carriers. This hypothesis can be tested by measuring the $T_e$ dependence of the mobilities when $T_e$ is varied by changing the electric field (see Fig. \ref{Fig02}(b)) while $T_L$ is held constant. In this fixed $T_L$, variable field measurement, the hole statistics do not change since $T_h = T_L$. The results of this measurement are also shown in Fig. \ref{Fig05}(b), the solid triangles corresponding to $T_L \approx 35$ K and the solid squares corresponding to $T_L = 15$ K. For 100 K $< T_e <$ 200 K the mobility dependence is well fitted by a $T_e^{-1.3}$ law, close to that obtained in the fixed field, variable $T_L$ case (circles). For $T_e < 50$ K there is a dramatic increase of mobility, with both electron and spin mobilities being described by a $T_e^{-4.3}$ power law.

In order to determine separately the temperature dependences of the two factors in Eq. \ref{tau}, $p$ can be measured using photoconductivity and photoHall measurements. Using a defocused laser excitation to ensure that the photoelectron concentration is homogeneous over the Hall bar, the ratio $r_H=\mu _e^H /\mu_e$ of the Hall mobility $\mu_e^H$ to the drift mobility of minority electrons can be found. $p$ can then be determined using \cite{popovic2004}  

\begin{equation}
r_H=\frac{\Gamma (5/2+2p)\Gamma (5/2)}{[\Gamma (5/2+p)]^2} 
\label{rh}
\end{equation}

The result is  $r_H = 0.95 \pm 0.25$ at $T_L =$ 15 K and $r_H = 0.8 \pm 0.25$ at $T_L =$ 300 K, in agreement with the values close to unity obtained for majority electrons in n-GaAs\cite{look1996}. Using these values and the graphical representation of Eq. \ref{rh} in the inset of Fig. \ref{Fig05}(b), $p$ is found to lie between approximately -0.5 and 0.5, in qualitative agreement with the predictions of the Brooks-Herring model \cite{chatto1981} for screened collisions by charged impurities. The small value of $p$ implies that the measured temperature dependences are mainly those of the prefactor $a(T)$ in Eq. \ref{tau}.

The origin of the $T_e$ dependence of the minority electron mobility is at present not understood. At the doping levels considered here it is probable that the mobility is determined by scattering off potential fluctuations caused by the random distribution of ionized acceptors and of their screening by valence holes. For $T_h =$ 10 K the maximum amplitude of the potential fluctuations is of the order of 40 meV \cite{efros1972} and while some attempts have been made to include the effects of screening on transport \cite{kaneto1993,chatto1981,quang1993}, a complete description including the effect of disorder is still out of reach. This work shows that in developing these models, accounting for the hole distribution alone \cite{kaneto1993} cannot describe the observed temperature dependence of the minority electron mobility since significant variations are measured even when the hole temperature is constant. The inclusion of minority carrier statistics in theoretical models is therefore necessary to better understand minority carrier mobilities in doped semiconductors.

\bibliographystyle{apsrev}


\end{document}